\documentclass[aps,twocolumn,amsmath,amsfonts,amssymb]{revtex4}
\usepackage{graphicx}

\begin{document}

\title{Polymers with self-attraction and stiffness: a generic phase structure}

\author{J.\ Krawczyk} \email{j.krawczyk@ms.unimelb.edu.au}
\author{A.\ L.\ Owczarek} \email{a.owczarek@ms.unimelb.edu.au} 
\author{T.\ Prellberg} \email{t.prellberg@qmul.ac.uk}
\affiliation{School of Mathematical Sciences, Queen Mary, University of London,
Mile End Road, London E1 4NS, United Kingdom}
\affiliation{Department of Mathematics and Statistics, The University of Melbourne, 
3010, Australia}

\begin{abstract}
  Recently it has been shown that a two-dimensional model of
  self-attracting polymers based on attracting segments displays two
  phase transitions, a $\theta$-like collapse between swollen polymers
  and a globular state and another between the globular state and a
  polymer crystal. On the other hand, the canonical model based on
  attracting monomers on lattice sites displays only one: the standard
  tricritical $\theta$ collapse transition.  Here we show that by
  considering both models with the addition of stiffness the two
  models display the \emph{same} generic phase diagram. In fact we
  claim that any two-dimensional model of a self-attracting single
  polymer in solution based upon a fully volume-excluded backbone with
  isotropic short range attraction should show this same phase
  structure.  We point to the model of hydrogen bonded polymers to
  demonstrate this observation. In three dimensions we note that more
  than one crystalline phase may occur.
  \end{abstract}

\maketitle

\vspace{-2ex}
\section{Introduction}
\vspace{-2ex}

The study of the thermodynamic behaviour of a single model molecule in
solution has be re-invigorated of late with a variety of new
experimental techniques able to tease out more refined features and
also with an escalating interest in biopolymers. It is therefore
perhaps  surprising to find that alternative simple models of a
single polymer display different phase structures: for some models
there exist only two thermodynamic phases, a high-temperature swollen
coil and a disordered liquid-like globule, whereas for other models
there is a third, compact phase with crystalline order.

In this letter we present a unified description of these models,
based on the study of a particular model that provides the missing link.
This model incorporates attractive interactions between segments of the 
polymer and bending stiffness. Computer simulations allow us to conclude 
that all the two-dimensional models in the literature are one or two parameter slices 
of a larger model in which only three phases exists.  We therefore \emph{conjecture}
that in any model of an isolated non-oriented polymer restricted to a
substrate (so being effectively two-dimensional), with homogeneous
short-ranged attractive interactions between parts of the polymer,
regardless of whether these parts are single monomers or groups of
monomers (so that an effective stiffness is introduced), only the three 
phases mentioned above will occur in the larger parameter space. 
%These three phases are a
%high temperature disordered swollen coil, a disordered liquid-like
%compact globule phase and an ordered compact crystalline phase. 
The phase transition between the swollen coil and the globule is described
by the traditional $\theta$-point, while the transition between the
swollen coil and the crystal is first order. Finally the most
interesting transition between the globule and the crystal is second
order (in two dimensions).

\vspace{-2ex}
\section*{Background}
\vspace{-2ex}

The canonical description of the thermodynamic states of a isolated
polymer has a reference high temperature disordered state known as the
`swollen coil' or `extended state' which is described by the $N
\rightarrow 0$ of the magnetic O($N$) model and correspondingly by the
$N \rightarrow 0$ of the $\varphi^4$ O($N$) field theory
\cite{gennes1972a-a}. The lattice model of self-avoiding walks is a
good model of this situation where in both two and three dimensions
the fractal dimension of the polymer $d_f$ is less than the dimension
of space and also less than the fractal dimension of simple random
walks (which is 2). The values of $d_f=1/\nu$, where $\nu$ is the
exponent describing the scaling of the radius of gyration of the
polymer with its length, are $4/3$ in two dimensions
\cite{nienhuis1982a-a} and $1.702(1)$ in three dimensions, have been
well studied \cite{li1995a-a}. It is expected that if the temperature
is lowered then the polymer will undergo a collapse at one particular
temperature, called the $\theta$-temperature. At temperatures lower than 
the $\theta$-temperature the fractal dimension of the polymer attains that
of space, that is $d_f=d$. The standard description of the collapse
transition is a tricritical point related to the $N \rightarrow 0$
limit of the $\varphi^4$--$\varphi^6$ O($N$) field theory
\cite{gennes1975a-a,stephen1975a-a,duplantier1982a-a}.
Thermodynamically this implies that there is a second-order phase
transition on lowering the temperature: the specific heat exponent
$\alpha$ is conjectured to be $-1/3$ in two dimensions
\cite{duplantier1987a-a} and $0$ in three dimensions with a
logarithmic divergence of the specific heat. In two dimensions the
fractal dimension of the polymer is expected to be $d_f=7/4$
\cite{duplantier1987a-a}. The low temperature state has been likened
to a liquid drop; it is compact but disordered
\cite{owczarek1993e-:a}. The corresponding lattice model which
displays this collapse $\theta$-transition is the \emph{Interacting
  Self-Avoiding Walk} model (ISAW) \cite{vanderzande1991a-a} where an
attractive potential is associated with pairs of sites of the walk
adjacent on the lattice though not consecutive along the walk.

However, more recently two other simple models of a single polymer have
been studied. These have been introduced in the context of biopolymers
where hydrogen bonding plays an important role \cite{pauling1951a-a},
and the interacting residua lie on  partially straight segments of
the chain. The first one, the \emph{Interacting Hydrogen-Bond} model (IHB)
\cite{bascle1993a-a,foster2001a-a,krawczyk2007b-:a}, modifies the ISAW
model such that a pair of sites on the self-avoiding walk acquires a
hydrogen-like bond potential if the sites are (non-consecutive) nearest
neighbours, as in the ISAW model, and each site lies on a straight
section of the walk. This model displays, in contrast to ISAW, a
single collapse transition which is \emph{first order} in both two and
three dimensions. Here the low temperature state is an anisotropic
ordered compact phase that is described as a polymer crystal. 

\begin{figure}[ht!]
  \centering
  \includegraphics[scale=0.7]{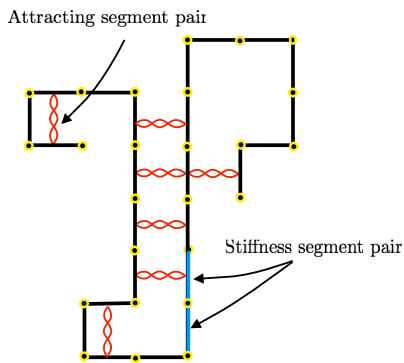}
  \caption{A self-avoiding walk with the interactions of the
    attracting segment (AS) model shown as intertwined curves between
    bonds of the walk on opposite sides of the squares of the
    lattice. Also shown is an example of a stiffness segment pair
    which obtains a stiffness energy in our generalisation.} 
  \label{fig_ssas-model}
\end{figure}

The second model introduced to account for hydrogen bonding is the
\emph{Attracting Segments} model
(AS) \cite{machado2001a-a,buzano2002a-a,foster2007a-a} (also known as
`interacting bonds'). It is a lattice model based upon self-avoiding walks
where an attractive potential is assigned to \emph{bonds} of the walk
that lie adjacent and parallel on the lattice (though not consecutive
along the walk), see Figure~\ref{fig_ssas-model}. This model has been
less extensively studied on regular lattices --- it has been studied
on the square lattice \cite{foster2007a-a} --- though \emph{two} phase
transitions have been identified, one of which is identified as the
$\theta$-point. 

%%%%%%%%%%%%%%%%%%%%%%%%%%
%%%%%%%%%%%%%%%%%%%%%%%%%%%

Crucially, there is a different modification of the ISAW model that displays two
phase transitions for a range of parameters, namely the
\emph{semi-flexible} ISAW model
\cite{bastolla1997a-a,vogel2007a-a,doye1997a-a}. Here two energies
are included: the nearest-neighbour site interaction of the ISAW model
and also a stiffness energy associated with consecutive parallel bonds
of the walk (equivalently, a bending energy for bends in the walk). This
has been studied on the cubic lattice by Bastolla and Grassberger
\cite{bastolla1997a-a}. 
They showed that
when there is a strong energetic preference for straight segments,
this model undergoes a single first-order transition from the
excluded-volume high-temperature state to a crystalline state.
On the other hand, if there is only a weak preference for straight
segments, the polymer undergoes two phase transitions. On lowering the
temperature the polymers undergoes a $\theta$-point transition to
the liquid globule followed at a lower temperature by a first-order
transition to the frozen crystalline phase. Recent work
\cite{krawczyk2008=a-:a} on the
semi-flexible ISAW model on the square lattice displays a similar
phase diagram though the  transition between the globule and the frozen state
has been seen to be second order. 

It would seem reasonable to conjecture that the phases and phase
transitions seen in the semi-flexible ISAW model are essentially those
seen in both the IHB model and the AS model. On the other hand the
addition of stiffness to the IHB model has been seen to change little
of the behaviour of that model \cite{krawczyk2008=a-:a}. Importantly,
the IHB model was recently extended \cite{krawczyk2007a-:a} to a
hybrid model (IHB-INH) that includes both the hydrogen-like bond
interactions and non-hydrogen like bond interactions, with separate
energy parameters. When the non-hydrogen bonding energy is set to zero
the IHB model is recovered.  If both energies are set to be the same
value then the ISAW \emph{without stiffness} is recovered. They found for
large values of the ratio of the interaction strength of
hydrogen-bonds to non-hydrogen bonds, that a polymer will undergo a single
first-order phase transition from a swollen coil at high temperatures
to a folded crystalline state at low temperatures.  On the other hand,
for any ratio of these interaction energies less than or equal to one
there is a single $\theta$-like transition from a swollen coil to a
liquid droplet-like globular phase. For intermediate
ratios two transitions can occur, so that the polymer first undergoes
a $\theta$-like transition on lowering the temperature, followed by a
second transition to the crystalline state. In three dimensions it was
found that this second transition is first order, while in two
dimensions they found that it is probably second order with a
divergent specific heat. In other words, by adding an energy to both
the hydrogen-like and non-hydrogen like interactions a phase diagram
similar to the one for the semi-flexible ISAW is found. We note that
the IHB model has been generalised on the cubic lattice in such a way
that two different crystalline phases were observed
\cite{krawczyk2007b-:a}. 

Here we bring together the pieces of this puzzle by studying the AS
model in the presence of stiffness on the square lattice. We
concentrate our study in two dimensions where only one type of
crystalline phase has been observed. 

\vspace{-2ex}
\section*{Semi-flexible Attracting Segments model}
\vspace{-2ex}

Our semi-flexible attracting segments model (semi-flexible AS model)
is a simple self-avoiding walk on square lattice, with
self-interactions as in the AS model
\cite{machado2001a-a,buzano2002a-a,foster2007a-a} and a stiffness (or
equivalently bend energy) added.  Specifically, the energy of a single
chain (walk) consists of two contributions (see
Figure~\ref{fig_ssas-model}): the energy $-\varepsilon_{as}$ for each
attracting segment pair, being a pair of occupied bonds of the lattice
that are adjacent and parallel on the lattice and not consecutive
along the walk; and an energy $-\varepsilon_{ss}$ for each stiffness
segment pair, being a pair of bonds consecutive along the walk that
are parallel. A walk configuration $\varphi_n$ of length $n$ has total
energy
\begin{equation}
E_n(\varphi_n) = -m_{as}(\varphi_n)\ \varepsilon_{as} - m_{ss}(\varphi_n)\ \varepsilon_{ss},
\end{equation}
where $m_{as}$ denote the number of attracting segment pairs and
$m_{ss}$ denotes the number of stiffness segment pairs.  The partition
function is defined then as
\begin{equation} Z_n(\beta_{as},\beta_{ss})=
\sum_{m_{as},m_{ss}} C_{n,m_{as},m_{ss}} e^{\beta_{as} m_{as}+\beta_{ss} m_{ss}},
\end{equation}
where $\beta_{as}= -\varepsilon_{as} /k_BT$ and $\beta_{ss}=
-\varepsilon_{ss}/k_BT$ for temperature $T$ and Boltzmann constant
$k_B$.  $C_{n,m_{as},m_{ss}}$ is the density of states, which we
have estimated by means of Monte Carlo simulations.

\vspace{-2ex}
\section{Results}
\vspace{-2ex}

On the square lattice we performed simulations using the
FlatPERM algorithm \cite{prellberg2004a-a}, estimating the density of
states up lengths for $n=128$ over the two parameters $m_{as}$ and
$m_{ss}$. We have also estimated the end-to-end distance as a measure
of the size of the polymer: this enables us to estimate the fractal
dimension of the polymer via estimation of the exponent $\nu$. The
density of states allows us to calculate the internal energy and the
specific heat, or equivalently, the mean values and the fluctuations
of $m_{as}$ and $m_{ss}$, respectively.
This allows to locate phase transitions through the possible
divergences in the specific heat.  To detect orientational order, we
estimated the \emph{anisotropy parameter} \cite{bastolla1997a-a}. In
two dimensions, denoting the number of bonds parallel to the $x$-,
and $y$-axes by $n_x$ and $n_y$, respectively, we define
\begin{equation}
\rho=1.0-\frac{\min(n_x,n_y)}{\max(n_x,n_y)}
\end{equation}
to be the anisotropy parameter.
In a system without orientational order, this quantity tends to zero
as the system size increases.  A non-zero limiting value less than one
of this quantity indicates weak orientational order with
$n_{min}\propto n_{max}$, while a limiting value of one indicates
strong orientational order, where $n_{max}\gg n_{min}$.

\begin{figure}[ht!] 
  \centering 
  \includegraphics[scale=0.7]{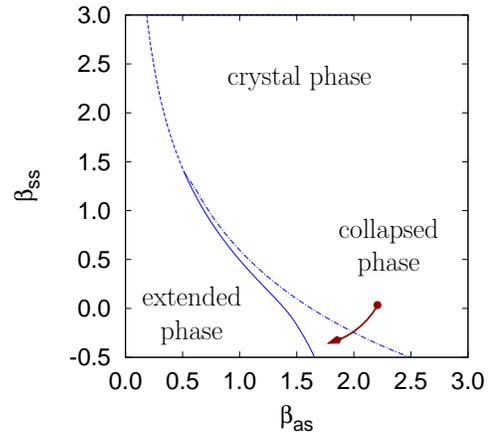} 
  \caption{Phase boundaries based upon data at length $128$.
The boundaries were 
found by looking for the maximum in fluctuation in the number of $as$
contacts (the left boundary) and the number of $ss$ 
segments (the right boundary). }  
  \label{fig_mod_asss} 
\end{figure}

Figure~\ref{fig_mod_asss} shows the peak positions of the
fluctuations in $m_{as}$ and $m_{ss}$ at length 128 from simulations on
the square lattice. It is clear from
shorter lengths that some of these peaks are asymptotically
diverging. We therefore have a finite-size approximation to a phase
diagram in this figure.
%%%%%%%%%%%%%%%%%%%%%%%%% 
%%%%%%%%%%%%%%%%%%%%%%%%%%% 

\begin{figure}[ht!]  
  \centering
  \includegraphics[scale=0.336]{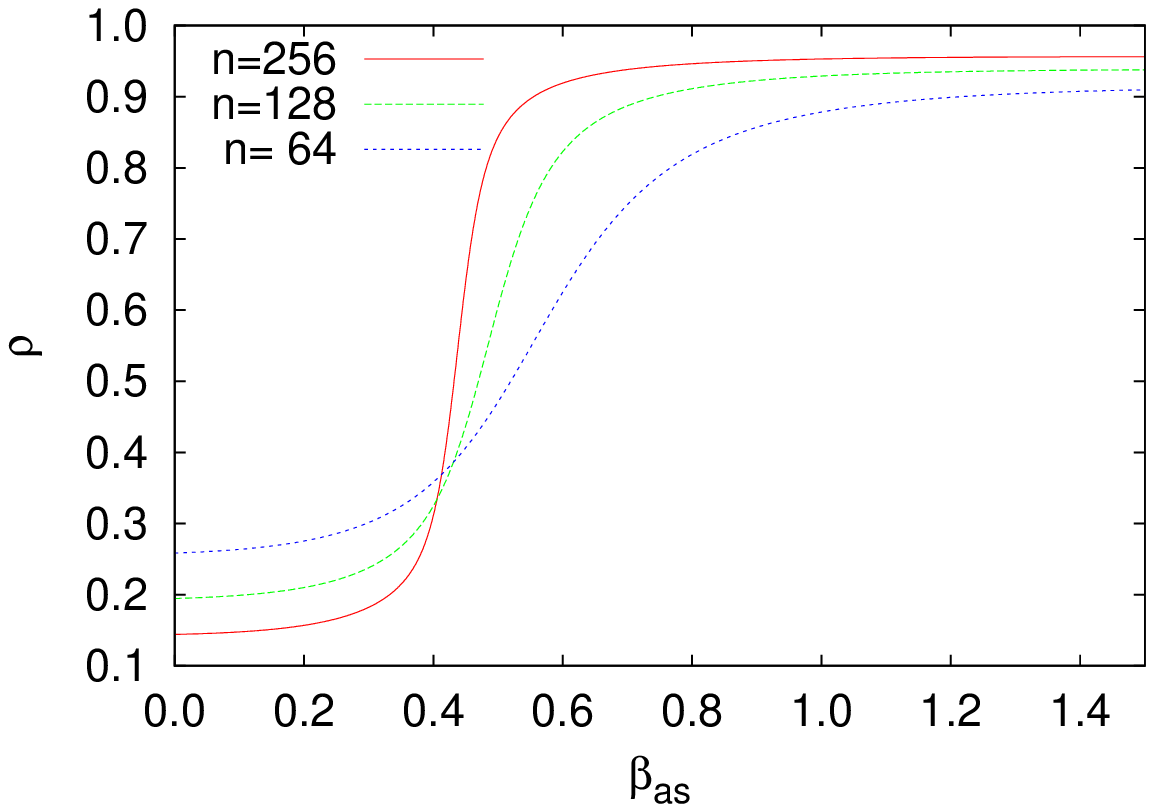}
\includegraphics[scale=0.336]{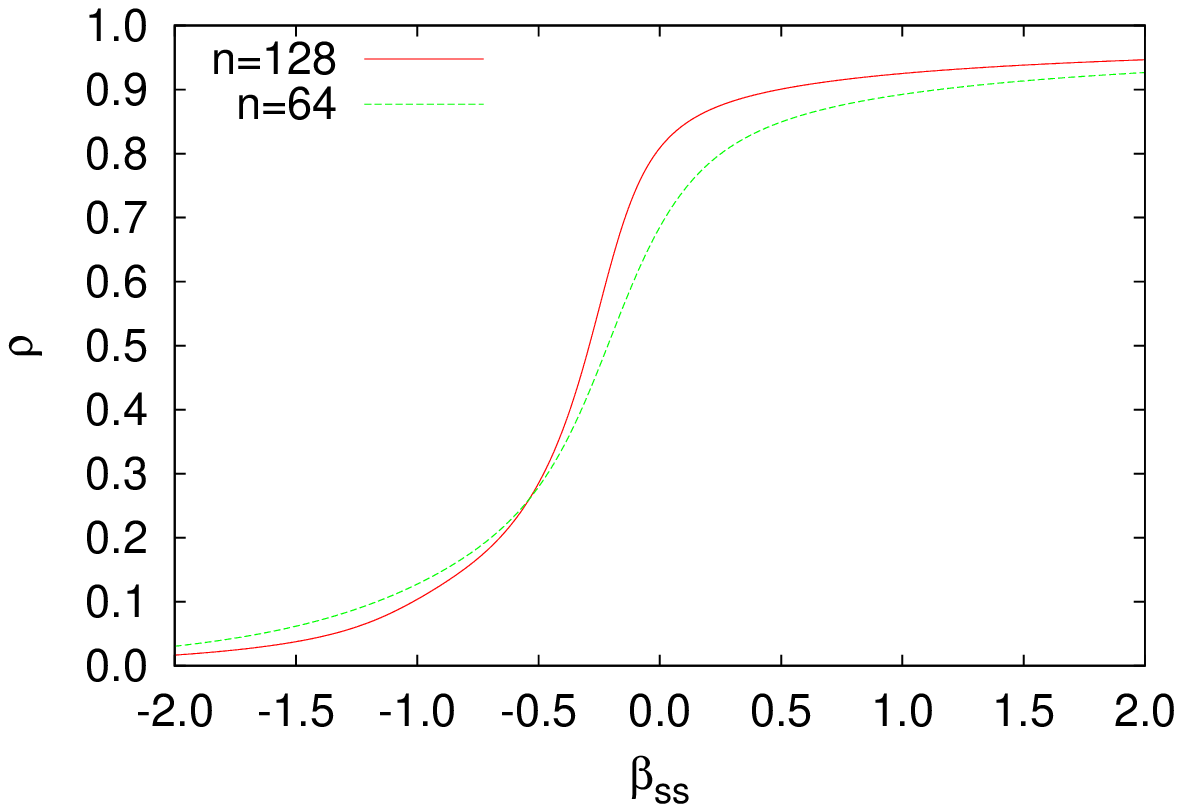}
  \caption{The anisotropy parameter $\rho$ for $\beta_{ss}=1.5$
    (left) and $\beta_{as}=2.0$ (right) for polymer lengths $n=64$,
    $128$ and $256$. For $\beta_{ss}=1.5$ it is converging to 0 for
    small $\beta_{as}$ and to 1 for large $\beta_{as}$, changing over
    roughly at $\beta_{as}=0.4$. For $\beta_{as}=2.0$ it is converging
    to 0 for small $\beta_{ss}$ and to 1 for large $\beta_{ss}$,
    changing over roughly at $\beta_{ss}=-0.5$.}
\label{fig_bs1.5_ro}  
\end{figure}

To characterise the possible phases in each region we have considered
the scaling of the end-to-end distance. For small $\beta_{as}$ we find
that the exponent $\nu$ is close to $3/4$: specifically we find
$d_f=1.33(1)$ for $(\beta_{as},\beta_{ss}) = (0.5,-0.5)$. On the other
side of the first curve separating small $\beta_{as}$ from larger
$\beta_{as}$ we find that $\nu$ is close to $1/2$ no matter how large
$\beta_{as}$ becomes: at points $(\beta_{as},\beta_{ss})$ being
$(1.8,-0.5)$, $(2.3,-0.5)$ and  $(3.0,-0.5)$ we find $d_f =2.00(1)$. Hence the two
phases to the right of the leftmost curve are dense. To
clarify the two low temperature phases we have measured the anisotropy
parameter. We find that the anisotropy parameter is tending to zero
for values of $\beta_{as}$ smaller that the right boundary and tending towards one for values of $\beta_{as}$ larger that the right boundary. That is, it tends to zero in the swollen and globular
phases and one in the crystal phase. In Figure~\ref{fig_bs1.5_ro} we show the anisotropy
parameter at fixed $\beta_{ss}=1.5$ and at fixed $\beta_{as}=2.0$
plotted for three different polymer lengths. 
There are stronger corrections to scaling in the globule to crystal
transition evident.
%%%%%%%%%%%%%%%%%%%%%%%%%%  
%%%%%%%%%%%%%%%%%%%%%%%%%%%  

\begin{figure}[h!]
  \centering
  \includegraphics[scale=0.6]{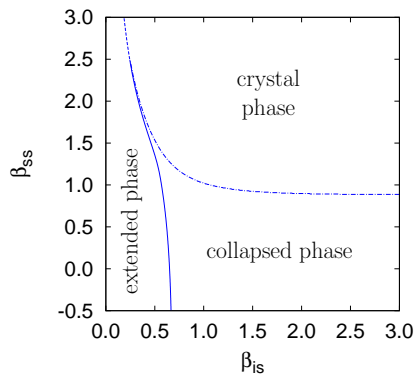}
  \caption{Finite-size phase diagram for the square lattice semi-flexible ISAW model. 
  }
  \label{fig_mod_isss}
\end{figure}

We therefore surmise that for high temperatures the polymer is in the
swollen phase, while at intermediate temperatures the polymer may be
in a globular state, and for low enough temperatures always enters the
anisotropic crystalline state. One can compare the finite size phase diagram that we have
to that of the semi-flexible ISAW model \cite{bastolla1997a-a,krawczyk2008=a-:a}, see
Figure~\ref{fig_mod_isss}. Apart from a shift of boundaries, so that for
$\beta_{ss}=0$ there is only one phase transition, the two diagrams are
remarkably similar.

To test the similarity of these diagrams further we have explored the
phase transition between the phases. It has been found that the
swollen-to-crystal transition in the semi-flexible ISAW model
\cite{bastolla1997a-a,krawczyk2008=a-:a} is first-order in two (and
three) dimensions. We find the same here for our semi-flexible AS
model in two dimensions: the scaling of fluctuations of $m_{as}$ have
a classic first-order behaviour with a linear divergence in polymer
length and a transition width scaling as the inverse length. The
globule-to-crystal transition has been found to be second order in two
dimensions.  To consider the globule-to-crystal transition on the
square lattice in our model more closely, we have also simulated
longer length polymers up to $n=512$ for constant $\beta_{ss}=-0.5$.
Using these simulations we find results similar to those found for the
semi-flexible ISAW model. The swollen-globule transition is expected
to be in the universality class of the $\theta$-point and its specific
heat signature is weak ($\alpha <0$). This is difficult to see at the
lengths considered here though the transition is clearly weakening
with length.

%%%%%%%%%%%%%%%%%%%%%%%%%%
%%%%%%%%%%%%%%%%%%%%%%%%%%%
We have therefore seen that the phase diagrams of the semi-flexible
ISAW model and our semi-flexible AS model contain the same phases and
phases transitions. In fact they have a very similar shape. 

Finally, it is instructive to consider the phase diagram of the extended
interacting hydrogen-bonding model \cite{krawczyk2007a-:a}.
While the different type of parameters make comparisons indirect, we
note that the same phases occur and same phase transitions as the two
semi-flexible models discussed in two dimensions. This supports the
idea that regardless of how stiffness is introduced in the interacting
polymer model, these three phases will occur in an appropriate
parameter space. To make this conclusion more concrete one can realise
that by enlarging the parameter space of concern in the manner
considered in \cite{buzano2002a-a} the ISAW model, the IHB and the AS
models are all specialisations of a more general model. Now we have
seen that with the addition of a further stiffness parameter the general phase
structure of the more general model should still only contain the three phases
and phases transition discussed above.

In this article we have developed a more complete and consistent
understanding of the phase structure of a single polymer in solution,
and restricted to a substrate so that it is effectively
two-dimensional, than had previously been available.  For the
three-dimensional model we expect a weaker result where more than one
crystalline phase may exist. (Our preliminary studies of the AS
model on the cubic lattice show that only one phase transition occurs which
makes it less relevant to the discussion here: the observations may be
understood as the weaker effect of AS model effective stiffness there.)
It should be in this context that experimental results of polymer collapse be analysed.

%\acknowledgments
%\vspace{-2ex}
 Financial support from the Australian Research Council and the Centre
 of Excellence for Mathematics and Statistics of Complex Systems is
 gratefully acknowledged by the authors.

\end{document}